\documentclass[12pt]{article}
\usepackage{amsmath,epsfig,graphicx,amssymb}
\usepackage[margin=0.7in]{geometry}
\newcommand{\beq}{\begin{equation}}
\newcommand{\eeq}{\end{equation}}
\newcommand{\ben}{\begin{eqnarray}}
\newcommand{\een}{\end{eqnarray}}
\newcommand{\bes}{\begin{subequations}}
\newcommand{\ees}{\end{subequations}}
\newcommand{\bFig}{\begin{figure}}
\newcommand{\eFig}{\end{figure}}

\date{}
\begin{document}

\title{Intensity-Intensity Correlations of Classically Entangled Light}
\author{Partha Ghose\footnote{partha.ghose@gmail.com} \\
Centre for Astroparticle Physics and Space Science (CAPSS),\\Bose Institute, \\ Block EN, Sector V, Salt Lake, Kolkata 700 091, India, \\and\\ Anirban Mukherjee \footnote{mukherjee.anirban.anirban@gmail.com}\\Indian Institute of Science Education \& Research,\\Mohanpur Campus, West Bengal 741252.}
\maketitle
\begin{abstract}
An experiment is proposed to show that after initial frequency and polarization selection, classical thermal light from two independent sources can be made path-polarization entangled. Such light will show new intensity-intensity correlations involving both path and polarization phases, formally similar to those for four-particle GHZ states. For fixed polarization phases, the correlations reduce to the Hanbury Brown-Twiss phase correlations. It is also shown that these classical correlations violate noncontextuality. 
\end{abstract}

\section{Introduction}
That entanglement, which has hitherto been regarded as an exclusively quantum feature of the world, also occurs in classical polarization optics, resulting in the violation of Bell-like inequalities, has been a major realization in recent years \cite{gm}. It has even been shown that entangled classical light violates noncontextuality \cite{gm2}. In this paper we extend the discussion to second-order interference effects and show first that light from two independent thermal sources, after initial frequency and polarization selection, can be made path-polarization entangled. It is then shown that the intensity-intensity correlations of such entangled classical light of variable polarization and path phases are formally similar to those of four-particle GHZ states \cite{ghz}, and lead to violation of noncontextuality. These correlations reduce to the familiar Hanbury Brown-Twiss correlations \cite{hbt} for fixed polarization.

Before proceeding further, we would like to clarify what we mean by states of classical light and nonquantum or classical entanglement. Coherent states are closest to classical light, and a polarized classical light beam (and not just specific modes) may be characterized by the expectation value of the electric field operator ${\bf \hat{E}}({\bf x},t)$ in a coherent state:
\ben
{\bf E}({\bf x},t) &=& \langle {\bf \hat{E}}({\bf x},t)\rangle = \sum_k \sqrt{2\pi\hbar \omega_k}\left[a_k(0) {\bf{\hat{e}}}_k e^{-i\omega_kt}u_k({\bf x}) + c c\right]\nonumber\\
&=&  {\bf \hat{e}} A ({\bf x},t)\label{sep}
\een
where $A ({\bf x},t)$ is the complex amplitude and the unit polarization vector ${\bf \hat{e}}$ is in a direction which is a linear combination of vertical and horizontal polarizations. This shows that a polarized light beam can be written as a tensor product $\vert A)\otimes \vert \lambda)\in {\cal{H}}_{path}\otimes {\cal{H}}_{pol}$ with $A({\bf x},t) \in {\cal{H}}_{path}$, the infinite dimensional Hilbert space of solutions of the scalar wave equation in optics, and $\vert \lambda) \in {\cal{H}}_{pol}$, the Hilbert space of polarization states, with 
\beq
\vert \lambda) = e^{i\phi}\left(\begin{array}{c}\cos\theta \\ e^{i\chi}\sin\theta\end{array}\right),
\eeq 
a linear combination of the transverse polarizations $\lambda_1$ and $\lambda_2$. This can also be written as the Jones vector \[\vert J) = \frac{1}{\sqrt{(J\vert J)}}\left(\begin{array}{c}
 E_x\\ E_y
\end{array} \right) \] with $E_x = \langle A_0\rangle \hat{e}_x {\rm exp (i\phi_x)}$ and $E_y = \langle A_0\rangle \hat{e}_y {\rm exp (i\phi_y)}$ the complex transverse electric fields, $\hat{e}_x$ and $\hat{e}_y$ the unit transverse polarization vectors, and $(J\vert J) = \vert E_x\vert^2 + \vert E_y\vert^2 =  \langle A_0\rangle^2 = I_0$, the intensity. (We prefer to use the notation $\vert X)$ to denote a classical light state to distinguish it from a quantum state $\vert X\rangle$.)

Ordinary classical thermal light is wholly unpolarized or partially polarized, and cannot be written in the separable product form (\ref{sep}). Such states must be written as
\beq
\vert E) = \vert h)\vert E_h) + \vert v)\vert E_v)
\eeq
where $\vert h)$ and $\vert v)$ are horizontal and vertical polarization states and $\vert E_h)$ and $\vert E_v)$ are function space optical fields. Such a state is clearly not separable in the path and polarization variables. More generally, an unpolarized beam (normalized to unit intensity) can be written as \cite{kian} 
\beq
\vert E) = \kappa_1 \vert u_1)\otimes \vert e_1) + \kappa_2 \vert u_2)\otimes \vert e_2)  \label{unpol}
\eeq
with $(u_j|u_k) = (e_j|e_k) = \delta_{jk}$ and $\kappa_1$ and $\kappa_2$ as normalization constants. The Schmidt theorem \cite{sch} guarantees such a decomposition. All these states are examples of classical nonseparability which is the essence of nonquantum or classical entanglement. 
 
Light sources such as lasers emit states of the optical field that are close to coherent states while thermal light sources emit a statistical mixture of coherent states with ensemble averages as intensities. Although light is considered to be fundamentally quantum in nature, there is an equivalence theorem \cite{sud} in optics that guarantees the complete equivalence of classical and quantum descriptions of all light except those that exhibit negative weights (quasi-probabilities) in the diagonal $P$ representation, such as Fock states and squeezed states. Thermal light is completely unpolarized with $\kappa_1=\kappa_2 = 1/\sqrt{2}$, and (\ref{unpol}) has the form of a Bell state. This makes it clear that entanglement, Bell states and violations of Bell-like inequalities are not exclusive to quantum mechanics. However, there is no nonlocality in classical optics. Nonlocality is a special feature of quantum entanglement associated with projective measurement and an ontic rather than epistemic interpretation of the wavefunction \cite{fuchs}.

\section{The Proposed Experiment}

\begin{figure}
\includegraphics[scale=0.7]{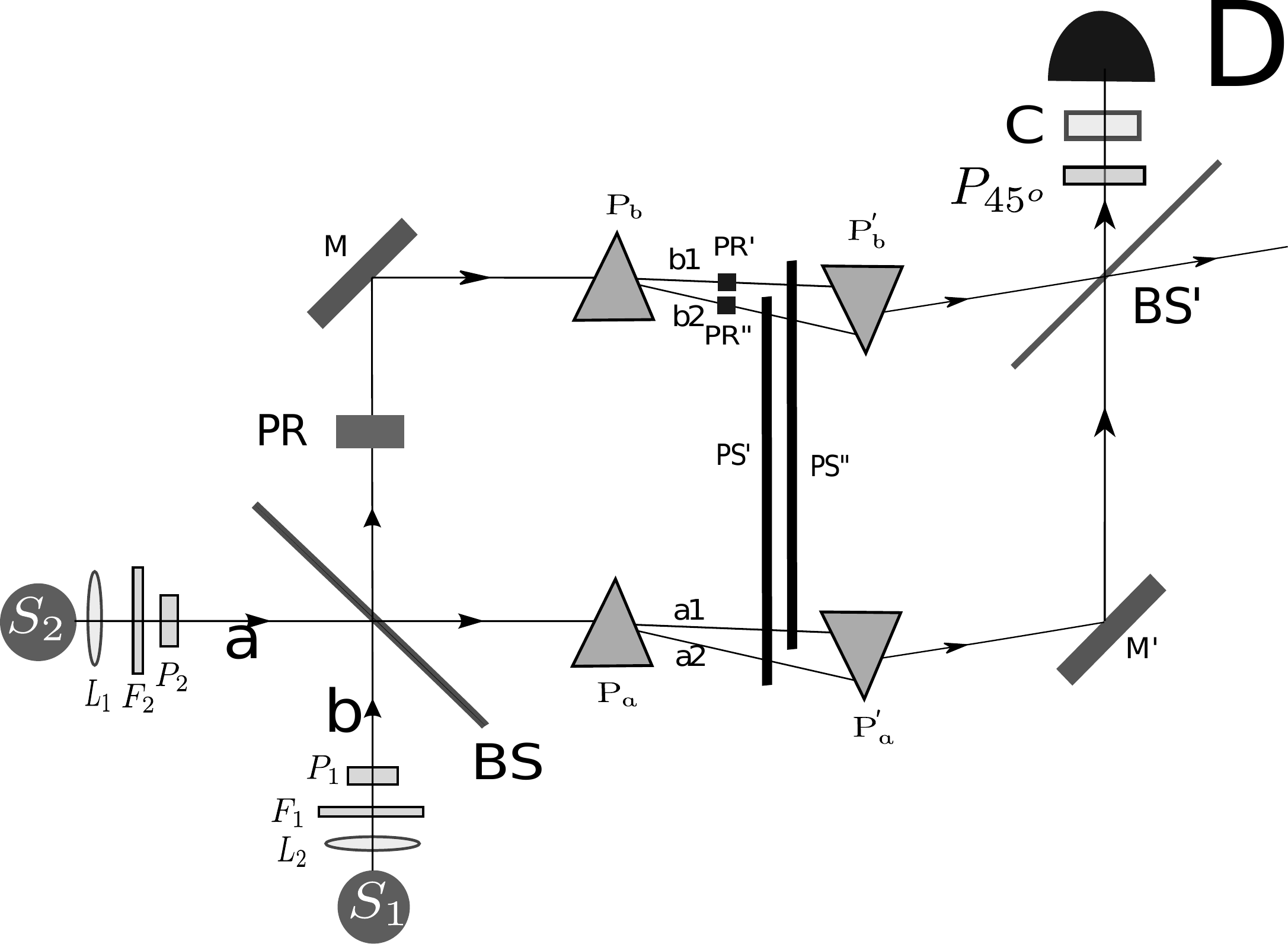}
\caption{$S_{1}$ and $S_{2}$ are two independent sources of thermal light, $L_1$ and $L_2$ are lenses that render the light plane parallel, $F_{1}$ and $F_{2}$ are two filters that select different frequencies $\omega_1$ and $\omega_2$, $P_1$ and $P_2$ are two polarization filters that select vertical polarization, $PR$ is a polarization rotator in path (b), $BS$ and $BS^{'}$ are two beam splitters, $M$ and $M^{'}$ are two mirrors, $P_{a}$ and $P_{a}^{'}$ are two prisms in path (a), $P_{b}$ and $P_{b}^{'}$ are two prisms in path (b), $PR^{'}$ and $PR^{''}$ are polarization rotators that can change the polarization of light, and $PS^{'}$, $PS^{''}$ are phase shifters for sources $S_{1}$ and $S_{2}$ respectively, $P_{45^{o}}$ is a polarization filter, and finally D is a detector.}  
\end{figure}
\subsection{Entanglement of Light from Two Thermal Sources}

Let $S_{1}$ and $S_{2}$ be two independent and incoherent sources of thermal light, $L_1$ and $L_2$ lenses that render the light plane parallel, and let the filters $F_{1}$ and $F_{2}$ select the frequencies $\omega_{1} \neq\omega_{2}$. Further, let $P_1$ and $P_2$ be two polarization filters that select vertical polarization. The states of the light beams from $S_{1}$ and $S_{2}$ incident on the beam splitter $BS$ can then be written as follows:
\begin{eqnarray}
\text{Source 1}:{\quad|\psi)_{S_{1}}} &=& A_{1}e^{i\omega_{1}t}|b)_{S_{1}} |V)_{S_{1}},\nonumber \\
\text{Source 2}:{\quad|\phi)_{S_{2}}} &=& A_{2}e^{i\omega_{2}t}|a)_{S_{2}} |V)_{S_{2}}
\end{eqnarray}
where $A_{1}$ and $A_{2}$ are the amplitudes of the electric field for the two sources with the filtered frequencies and polarizations respectively.
The beam splitter $BS$ splits each light beam light into two paths, and this process does not depend on the frequency of light. The unitary matrix describing this process is given by
\begin{eqnarray}
\hat{U}_{BS}=\frac{1}{\sqrt{2}}\left[|a)( a|-|b)( b|+|a)( b|+|b)( a|\right].
\end{eqnarray}
The states of light after $BS$ are therefore
\begin{eqnarray}
\text{Source 1}:\quad|\psi)_{S_{1}} &=& e^{i\omega_{1}t}\frac{A_{1}}{\sqrt{2}}\left(|a)_{S_{1}}+|b)_{S_{1}}\right) |V)_{S_{1}},\nonumber \\
\text{Source 2}:\quad|\phi)_{S_{2}} &=& e^{i\omega_{2}t}\frac{A_{2}}{\sqrt{2}}\left(|a)_{S_{2}}-|b)_{S_{2}}\right)|V)_{S_{2}}. 
\end{eqnarray}
The light in path $b$ then passes the polarization rotator $PR$, and hence, with
\begin{eqnarray}
\hat{U}_{PR}&=&|H)( V|+|V)(H|,\nonumber
\een
the states become
\ben
\text{Source 1}:\quad|\psi)_{S_{1}} &=&e^{i\omega_{1}t}\left(|a)_{S_{1}}|V)_{S_{1}}+|b)_{S_{1}}|H)_{S_{1}}\right), \nonumber \\
\text{Source 2}:\quad|\phi)_{S_{2}} &=&e^{i\omega_{2}t}\left(-|a)_{S_{2}}|V)_{S_{2}}+|b)_{S_{2}}\right |H)_{S_{2}}). 
\end{eqnarray}
Since the two light sources are independent,  one would expect the complete state of light to be a direct product state of the two light beams. However, since the two sources overlap as observed from both the paths (a) and (b) after $BS$, the complete state must be symmetrized as follows:
\begin{eqnarray}
|\Psi^{0}) &=&\frac{1}{\sqrt{2}}\left[|\psi)_{S_{1}}|\phi)_{S_{2}}+ |\phi)_{S_{1}}|\psi)_{S_{2}}\right]\nonumber\\
			 &=&e^{i(\omega_{1}+\omega_{2})t}\frac{A_{1}A_{2}}{\sqrt{2}}\left[|a)_{S_{1}}|V)_{S_{1}}|a)_{S_{2}}|V)_{S_{2}}-|b)_{S_{1}}|H)_{S_{1}}|b)_{S_{2}}|H )_{S_{2}}\right]. \label{sym}
\end{eqnarray}
The state then passes through the various optical elements as shown in Figure 1 and, before entering the second beam splitter $BS^{'}$, it evolves to (see Appendix-1)
\begin{eqnarray}
|\Psi)&=&\frac{A_{1}A_{2}}{\sqrt{2}}\left[|a)_{S_{1}}|V)_{S_{1}}|a)_{S_{2}}|V)_{S_{2}}-e^{i\left(\theta_{1}+\phi_{1}-\theta_{2}-\phi_{2}\right)}|b)_{S_{1}}|H)_{S_{1}}|b)_{S_{2}}|H )_{S_{2}}\right].\label{prestate}
\end{eqnarray}
After $BS^{'}$ the symmetrized state becomes 
\begin{eqnarray}
|\Psi^{f})&=&\hat{U}_{BS^{'}}^{S_{1}}\otimes\mathbb{I}^{S_{1}}_{pol}\otimes \hat{U}_{BS^{'}}^{S_{2}}\otimes\mathbb{I}^{S_{2}}_{pol}|\Psi) \label{poststate}\\
 &=&\frac{A_{1}A_{2}}{\sqrt{2}}\frac{|a)_{S_{1}}+|b)_{S_{1}}}{\sqrt{2}}|V)_{S_{1}}\frac{|a)_{S_{2}}+|b)_{S_{2}}}{\sqrt{2}}|V)_{S_{2}}\nonumber\\ &-&e^{i\left(\theta_{1}+\phi_{1}-\theta_{2}-\phi_{2}\right)}\frac{A_{1}A_{2}}{\sqrt{2}}\frac{|a)_{S_{1}}-|b)_{S_{1}}}{\sqrt{2}}|H)_{S_{1}}\frac{|a)_{S_{2}}-|b)_{S_{2}}}{\sqrt{2}}|H )_{S_{2}}.\label{HBTcorr}
\end{eqnarray}

\subsection{New Intensity-Intensity Correlations}
Let us consider the two observables 
\begin{eqnarray}
\sigma_{1\theta_{1}}^{pol}=e^{i\theta_{1}}|H)_{S_{1}}( V|_{S_{1}}+e^{-i\theta_{1}}|V)_{S_{1}}( H|_{S_{1}},\\
\sigma_{2\theta_{2}}^{pol}=e^{-i\theta_{2}}|H)_{S_{2}}( V|_{S_{2}}+e^{i\theta_{2}}|V)_{S_{2}}( H|_{S_{2}},
\end{eqnarray}  
operating on the polarization Hilbert spaces of the lights from the sources $S_{1}$ and $S_{2}$,
as well as the two observables 
\begin{eqnarray}
\sigma_{1\phi_{1}}^{path}&=&e^{-i\phi_{1}}|a)_{S_{1}}( b|_{S_{1}}+e^{i\phi_{1}}|b)_{S_{1}}( a|_{S_{1}},\\
\sigma_{2\phi_{2}}^{path}&=&e^{i\phi_{2}}|a)_{S_{2}}( b|_{S_{2}}+e^{-i\phi_{2}}|b)_{S_{2}}( a|_{S_{2}}
\end{eqnarray}
operating on the path Hilbert spaces. Let us define the normalized correlation
\begin{equation}
C(\theta_{1},\phi_{1}, \theta_{2},\phi_{2})=\frac{(\Psi^{o}|\sigma^{path}_{1\phi_{1}}.\sigma^{pol}_{1\theta_{1}}\otimes\sigma^{path}_{2\phi_{2}}.\sigma^{pol}_{2\theta_{2}}|\Psi^{o})}{(|A_{1}|^{2}+|A_{2}|^{2})^{2}} \label{cor}
\end{equation}
for the symmetrized state $|\Psi^{o})$ given by Eqn. (\ref{sym}). This can be written as a linear combination of intensity-intensity correlations, as we will now show. First, it can be shown (see Appendix-2 for details) that this correlation can be expressed in terms of the final state $|\Psi^{f})$ as
\begin{eqnarray}
C(\theta_{1},\phi_{1},\theta_{2},\phi_{2})&=&\sum_{(k,l,m,n)=0}^{1}(-1)^{k+l+m+n}(\Psi^{f} |\hat{I}_{S_{1}}(k\pi ,l\pi). \hat{I}_{S_{2}}(m\pi,n\pi)|\Psi^{f})\nonumber\\
&=&\frac{4|A_{1}|^{2}|A_{2}|^{2}}{\left(|A_{1}|^{2}+|A_{2}|^{2}\right)^{2}}\cos(\theta_{1}-\theta_{2}+\phi_{1}-\phi_{2}).\label{14}
\end{eqnarray}
where 
\beq
(\Psi^{f} |\hat{I}_{S_{1}}(k\pi ,l\pi). \hat{I}_{S_{2}}(m\pi,n\pi)|\Psi^{f})=\frac{2|A_{1}|^{2}|A_{2}|^{2}}{\left(|A_{1}|^{2}+|A_{2}|^{2}\right)^{2}}\left[1-\cos\left(\theta_{1}-\theta_{2}+[k-m]\pi+\phi_{1} -\phi_{2}+[l-n]\pi\right)\right].\label{15}
\eeq
Now, the Hanbury Brown-Twiss intensity-intensity correlations $g^{(2)}(\alpha, \beta)$ for non-entangled light ($\alpha,\beta$ being path differences) are known to have the form
\begin{eqnarray}
g^{(2)}(\alpha ,\beta)&=&\frac{|A_{1}|^{4}}{(|A_{1}|^{2}+|A_{2}|^{2})^{2}}+\frac{|A_{2}|^{4}}{\left(|A_{1}|^{2}+|A_{2}|\right)^{2}}+ \frac{2|A_{1}|^{2}|A_{2}|^{2}}{(|A_{1}|^{2}+|A_{2}|^{2})^{2}}\left(1-\cos(\alpha -\beta)\right) 
\end{eqnarray}
The first two terms correspond to two distinguishable processes, namely light from source $S_{1}$ going to the detector $D$ via paths $a$ and $b$ and similarly for the source $S_{2}$. The third term is the interference effect. 

For path-polarization entangled states one can define the correlation
\begin{eqnarray}
g^{(2)}(\theta_{1}-\theta_{2}+[k-m]\pi ,\phi_{1}-\phi_{2}+[l-n]\pi)&=& \frac{|A_{1}|^{4}}{(|A_{1}|^{2}+|A_{2}|^{2})^{2}} +\frac{|A_{2}|^{4}}{\left(|A_{1}|^{2}+|A_{2}|\right)^{2}}\nonumber\\&+&(\Psi^{f} |\hat{I}_{S_{1}}(k\pi ,l\pi). \hat{I}_{S_{2}}(m\pi,n\pi)|\Psi^{f}).
\end{eqnarray}
Using Eqn. (\ref{15}), we get
\beq
g^{(2)}(\theta_{1}-\theta_{2}+[k-m]\pi ,\phi_{1}-\phi_{2}+[l-n]\pi)= 1-\frac{2|A_{1}|^{2}|A_{2}|^{2}}{(|A_{1}|^{2}+|A_{2}|^{2})^{2}}\cos\left(\theta_{1}-\theta_{2}+[k-m]\pi+\phi_{1} -\phi_{2}+[l-n]\pi\right)
\eeq
Therefore, the correlation (\ref{14}) can be written as a linear combination of generalized Hanbury Brown-Twiss correlations for different sets of path and polarization differences: 
\begin{eqnarray}
C(\theta_{1},\phi_{1},\theta_{2},\phi_{2})=\sum_{(k,l,m,n)=0}^{1}(-1)^{k+l+m+n} g^{(2)}(\theta_{1}-\theta_{2}+[k-m]\pi ,\phi_{1}-\phi_{2}+[l-n]\pi).
\end{eqnarray}
From now on we will take $|A_{1}|^{2}$=$|A_{2}|^{2}$, i.e. the intensities of the two sources to be equal. For this special case 
\begin{eqnarray}
g^{(2)}(\theta_{1}-\theta_{2},\phi_{1}-\phi_{2})&=&1-\frac{1}{2}\cos\left(\theta_{1}-\theta_{2}+\phi_{1}-\phi_{2}\right),\nonumber\\
C&=&\cos\left(\theta_{1}-\theta_{2}+\phi_{1}-\phi_{2}\right).\label{ghz}
\end{eqnarray}
The forms of $\hat{I}_{S_{1}}(k\pi ,l\pi)$ and $\hat{I}_{S_{2}}(m\pi ,n\pi)$ are given by Eqns.(\ref{Intensity}) in Appendix-2.

To measure this cross-correlation it is sufficient to use a single detector $D$ placed after a $45^{o}$ polarizer $P_{45^{o}}$ and a second-harmonic-generation crystal. Since the output from $BS^{'}$ consists of two coincident beams from the two sources, the total intensity auto-correlation $A$ is given by
\ben
A&=&\int_{-\infty}^{\infty} |\vec{E}_{S_{1}}+\vec{E}_{S_{2}}|^{4}dt\nonumber\\
&=&\int_{-\infty}^{\infty}I^{2}_{S_{1}}(t) dt + \int_{-\infty}^{\infty}I^{2}_{S_{2}}(t) dt + 2\int_{-\infty}^{\infty}I_{S_{1}}(t)I_{S_{2}}(t)dt .
\een 
The first order interference term averages out to zero, and the third term corresponds to the generalized Hanbury Brown-Twiss correlation for the set ($\theta_{1}+k\pi,\theta_{2}+m\pi,\phi_{1}+l\pi,\phi_{2}+n\pi$). 

Eqn. (\ref{ghz}) shows that if three of the phase angles are measured, the fourth angle can be predicted. This is formally identical to the case of 4-particle GHZ states \cite{ghz}, the four quantum particles being replaced by the four classical light beams $a1, a2, b1, b2$  with tunable path and polarizations. However, there is an important difference. Note that this correlation will have the maximum value $+1$ and the minimum value $-1$ when
\begin{eqnarray}
\theta_{1}+\phi_{1}-\theta_{2}-\phi_{2}=0,\pi
\end{eqnarray}
respectively. In the quantum GHZ case, if three of these phase angles are measured, the fourth angle can be predicted with certainty only in these two cases. 

If the polarization phase angles $\theta_1, \theta_2$ are kept fixed, the correlation reduces to the familiar Hanbury Brown-Twiss (HBT) correlation.
 In a typical HBT set up two detectors are used and the distance between them is varied to introduce a variable phase difference between the two light paths. In our set up (Fig. 1) all phase differences can be varied continuously by using the optical elements $PR^{'}, PR^{''}, PS{'}, PS^{''}$, and the polarization filter $P_{45^{o}}$ projects the light into one of the basis states, as shown in Appendix-2. Hence, a single detector suffices.
 
\section{Violation of Noncontextuality}
{\flushleft{{\em Case 1}}}
Let $\phi=\phi_{1}-\phi_{2}$ and $\theta=\theta_{1}-\theta_{2}$. These can be tuned using the polarization rotators and path phase shifters. Writing the correlation function C as
\begin{eqnarray}
\overline{C}(\theta,\phi)=\cos(\theta+\phi),\label{cor2}
\end{eqnarray}
one can see that these two relative phases of the two independent light sources are contextually related, i.e. the measurement of one influences the measurement of the other although they belong to disjoint Hilbert spaces and are therefore expected to be noncontextual variables.

One can also define a quantity 
\begin{eqnarray}
S(\theta ,\phi ;\theta^{'} ,\phi^{'})= \overline{C}(\theta ,\phi) + \overline{C}(\theta ,\phi^{'})- \overline{C}(\theta^{'} ,\phi)+\overline{C}(\theta^{'} ,\phi^{'})
\end{eqnarray} 
and observe that the bound $-2\leq S\leq 2$, imposed by the requirement that the value of each of the four terms, though continuously variable and non-discrete, must lie between $\pm 1$ because of (\ref{cor2}), is clearly violated for the set $\theta=0, \theta^{'}=\pi/2, \phi=\pi/4, \phi^{'}=-\pi/4$ for which $S=2\sqrt{2}$. 

{\flushleft{{\em Case 2}}}

One can also fix the polarization phase of the source $S_{2}$ to be $\theta_{2}=\alpha$ and the path phase of the source $S_{1}$ to be $\phi_{1}=\beta$ such that $\alpha -\beta = 0$. Then the correlation function can be expressed as 
\begin{eqnarray}
C(\theta_{1}, \alpha ,\beta ,\phi_{2})=\tilde{C}(\theta_{1},\phi_{2})=\cos(\theta_{1}-\phi_{2}). \label{cor3}
\end{eqnarray}
Hence, the path phase of the source $S_{1}$ is related to the polarization phase of the source $S_{2}$ although they belong to disjoint Hilbert spaces and are expected to be mutually independent and noncontextual variables.

One can again define an $S^{'}$ function
\begin{eqnarray}
S^{'}(\theta_{1},\phi_{2};\theta^{'}_{1},\phi^{'}_{2})=\tilde{C}(\theta_{1},\phi_{2})+ \tilde{C}(\theta_{1},\phi^{'}_{2})-\tilde{C}(\theta^{'}_{1},\phi_{1})+\tilde{C}(\theta^{'}_{1},\phi^{'}_{2})
\end{eqnarray}
and observe that the bound $-2\leq S^{'}\leq 2$ imposed by (\ref{cor3}) is violated for the set $\theta_{1}=\alpha, \theta_{1}^{'}=\pi/2+\alpha , \phi_{2}=\beta-\pi/4, \phi_{2}^{'}=\beta+\pi/4$ for which $S^{'}=2\sqrt{2}$. 

Similarly, one can show that the path phase $\phi_2$ of the source $S_{2}$ is contextually related to the polarization phase $\theta_1$ of the source $S_{1}$ although they belong to disjoint Hilbert spaces and are expected to be mutually independent.

\section{Concluding Remarks}
That classical polarization optics, the paradigm of classical physics, shares many common features with quantum mechanics including entanglement and contextuality has its origin in their common Hilbert space structure. There are, admittedly, some profound differences between them, originating in the special nature of measurement in quantum theory. These have been discussed in Refs. \cite{gm}. This relatively recent realization opens up many new possibilities in classical optics that were inconceivable a few years ago. This paper explores one of them, namely in the area of intensity-intensity cross-correlations of two independent sources of thermal light. It is a generalization of the famous Hanbury Brown-Twiss effect to the case where both the path and polarization degrees of freedom of the light beams can be varied independently to produce new states of light formally analogous to 4-particle GHZ states in which these degrees of freedom lose their classical realism and become contextual.

\section{Acknowledgement}
The authors are grateful to C.S. Unnikrishnan for helpful comments on the first draft of this paper which have led to the addition of cmany clarifications. PG thanks the National Academy of Sciences, India for the award of a Senior Scientist
Platinum Jubilee Fellowship which allowed this work to be undertaken.

\section{Appendix-1}
Consider Figure 1. Light disperses into two paths at the prisms in both paths a and b. The polarization rotator $PR^{''}$ in path $b2 = b+\epsilon(\omega_{2})$ produces a change of the polarization phase of the light in that path from $|H)_{S_{2}}\rightarrow e^{-i\theta_{2}}|H)_{S_{2}}$. Similarly, the polarization rotator $PR^{'}$ in path $b1=b+\epsilon(\omega_{1})$ produces a change of the polarization phase of the light in that path from $|H)_{S_{1}}\rightarrow e^{i\theta_{1}}|H)_{S_{1}}$. Therefore, the states get modified to 
\begin{eqnarray}
\quad|\psi)_{S_{1}} &\rightarrow& e^{i\omega_{1}t}\frac{A_{1}}{\sqrt{2}}\left[ |a+\epsilon(\omega_{1}))_{S_{1}}|V)_{S_{1}} +e^{i\theta_{1}}|b+\epsilon(\omega_{1}))_{S_{1}}|H)_{S_{1}}\right], \\
\quad|\phi)_{S_{2}} &\rightarrow& e^{i\omega_{2}t}\frac{A_{2}}{\sqrt{2}}\left[-|a+\epsilon(\omega_{2}))_{S_{2}}|V)_{S_{2}} + e^{-i\theta_{2}}|b+\epsilon(\omega_{2}))_{S_{2}}|H)_{S_{2}}\right].
\end{eqnarray}
Then the light passes through the path phase shifter $PS^{'}$ across the paths $b1$ and $a1$ to produce a change of path phase $|b+\epsilon(\omega_{1}))_{S_{1}}\rightarrow e^{i\phi_{1}}|b+\epsilon(\omega_{1}))_{S_{1}}$ of the light from the source $S_1$. Similarly, the path phase shifter $PS^{''}$ across the patha $b_{2}$ and $a_{2}$ produces a change of path phase $|b+\epsilon(\omega_{2}))_{S_{2}}\rightarrow e^{-i\phi_{2}}|b+\epsilon(\omega_{2}))_{S_{2}}$ of the light from the source $S_2$.
Consequently, they modify the states to 
\ben
\quad |\psi)_{S_{1}} \rightarrow e^{i\omega_{1}t}\frac{A_{1}}{\sqrt{2}}\left[|a+\epsilon(\omega_{1}))_{S_{1}}|H)_{S_{1}} + e^{i(\theta_{1}+\phi_{1})}|b+\epsilon(\omega_{1}))_{S_{1}}|V)_{S_{1}}\right],\\ 
\quad|\phi)_{S_{2}} \rightarrow e^{i\omega_{2}t}\frac{A_{2}}{\sqrt{2}}\left[-|a+\epsilon(\omega_{2}))_{S_{2}}|H)_{S_{2}} + e^{-i(\theta_{2}+\phi_{2})}|b+\epsilon(\omega_{2}))_{S_{2}}|H)_{S_{2}}\right].
\een
Finally, the actions of the inverted prisms $P'_{a}, P'_{b}$ are given by the unitary matrices $\hat{U}^{S_{1}}_{P'_{b}P'_{a}}=\hat{U^{\dagger}}^{S_{1}}_{P_{b}P_{a}}$ and $\hat{U}^{S_{2}}_{P'_{b}P'_{a}}=\hat{U^{\dagger}}^{S_{2}}_{P_{b}P_{a}}$, and the states finally become
\begin{eqnarray}
\quad|\psi)_{S_{1}}\rightarrow e^{i\omega_{1}t}\frac{A_{1}}{\sqrt{2}}\left[|a)_{S_{1}}|H)_{S_{1}} +e^{i(\theta_{1}+\phi_{1})}|b)_{S_{1}}|V)_{S_{1}}\right],\\
\quad|\phi)_{S_{2}}\rightarrow e^{i\omega_{2}t}\frac{A_{2}}{\sqrt{2}}\left[-|a)_{S_{2}}|H)_{S_{2}} +e^{-i(\theta_{2}+\phi_{2})}|b)_{S_{2}}|V)_{S_{2}}\right].
\end{eqnarray}
Hence, the light evolves to the symmetrized state
\begin{eqnarray}
|\Psi)&=&\frac{A_{1}A_{2}}{\sqrt{2}}\left[|a)_{S_{1}}|V)_{S_{1}}|a)_{S_{2}}|V)_{S_{2}}-e^{i\left(\theta_{1}+\phi_{1}-\theta_{2}-\phi_{2}\right)}|b)_{S_{1}}|H)_{S_{1}}|b)_{S_{2}}|H )_{S_{2}}\right]
\end{eqnarray}
before entering the second beam splitter $BS^{'}$ whose action is described by the unitary operator
\begin{eqnarray}
\hat{U}_{BS^{'}}=\frac{1}{\sqrt{2}}\left[|a)( a|-|b)( b|+|a)( b|+|b)( a|\right].
\end{eqnarray}
After $BS^{'}$ the symmetrized state is given by 
\begin{eqnarray}
|\Psi^{f})&=&\hat{U}_{BS^{'}}^{S_{1}}\otimes\mathbb{I}^{S_{1}}_{pol}\otimes \hat{U}_{BS^{'}}^{S_{2}}\otimes\mathbb{I}^{S_{2}}_{pol}|\Psi)\\
 &=&\frac{A_{1}A_{2}}{\sqrt{2}}\frac{|a)_{S_{1}}+|b)_{S_{1}}}{\sqrt{2}}|V)_{S_{1}}\frac{|a)_{S_{2}}+|b)_{S_{2}}}{\sqrt{2}}|V)_{S_{2}}\nonumber\\ &-&e^{i\left(\theta_{1}+\phi_{1}-\theta_{2}-\phi_{2}\right)}\frac{A_{1}A_{2}}{\sqrt{2}}\frac{|a)_{S_{1}}-|b)_{S_{1}}}{\sqrt{2}}|H)_{S_{1}}\frac{|a)_{S_{2}}-|b)_{S_{2}}}{\sqrt{2}}|H )_{S_{2}}.
\end{eqnarray}

\section{Appendix-2}
Consider the observables
\begin{eqnarray}
\sigma^{pol}_{1\theta_{1}}=\sigma^{pol}_{1\theta_{1},0}-\sigma^{pol}_{1\theta_{1},\pi},\\
\sigma^{pol}_{2\theta_{2}}=\sigma^{pol}_{2\theta_{2},0}-\sigma^{pol}_{1\theta_{2},\pi},\\
\sigma^{path}_{1\phi_{1}}=\sigma^{path}_{1\phi_{1},0}-\sigma^{path}_{1\phi_{1},\pi},\\
\sigma^{path}_{2\phi_{2}}=\sigma^{path}_{2\phi_{2},0}-\sigma^{path}_{1\phi_{2},\pi},
\end{eqnarray}
with
\begin{eqnarray}
\sigma^{pol}_{1\theta_1 ,0}&=&I^{path}_{1}\otimes\frac{1}{2}(|V)_{S_{1}}+e^{i\theta_{1}}|H)_{S_{1}})((V|_{S_{1}}+e^{-i\theta_{1}}( H|_{S_{1}}),\nonumber \\
\sigma^{pol}_{1\theta_1 ,\pi}&=&I^{path}_{1}\otimes\frac{1}{2}(|V)_{S_{1}}-e^{i\theta_{1}}|H)_{S_{1}})(( V|_{S_{1}}-e^{-i\theta_{1}}( H|_{S_{1}}),\nonumber \\
\sigma^{path}_{1\phi_1 ,0}&=&\frac{1}{2}(|a)_{S_{1}}+e^{i\phi_{1}}|b)_{S_{1}})(( a|_{S_{1}}+e^{-i\phi_{1}}( b|_{S_{1}})\otimes I^{pol}_{1},\nonumber \\
\sigma^{path}_{1\phi_1 ,\pi}&=&\frac{1}{2}(|a)_{S_{1}}-e^{i\phi_{1}}|b)_{S_{1}})(( a|_{S_{1}}-e^{-i\phi_{1}}( b|_{S_{1}})\otimes I^{pol}_{1},\label{S1}
\end{eqnarray}
and
\begin{eqnarray}
\sigma^{pol}_{2\theta_2 ,0}&=&I^{path}_{2}\otimes\frac{1}{2}(|V)_{S_{2}}+e^{-i\theta_{2}}|H)_{S_{2}})((V|_{S_{2}}+e^{i\theta_{2}}( H|_{S_{2}}),\nonumber\\
\sigma^{pol}_{2\theta_2 ,\pi}&=&I^{path}_{2}\otimes \frac{1}{2}(|V)_{S_{2}}-e^{-i\theta_{2}}|H)_{S_{2}})(( V|_{S_{2}}-e^{i\theta_{2}}( H|_{S_{2}}),\nonumber \\
\sigma^{path}_{2\phi_2 ,0}&=&\frac{1}{2}(|a)_{S_{2}}+e^{-i\phi_{2}}|b)_{S_{2}})(( a|_{S_{2}}+e^{i\phi_{2}}( b|_{S_{2}})\otimes I^{pol}_{2},\nonumber \\
\sigma^{path}_{2\phi_2 ,\pi}&=&\frac{1}{2}(|a)_{S_{2}}-e^{-i\phi_{2}}|b)_{S_{2}})(( a|_{S_{2}}-e^{i\phi_{2}}( b|_{S_{2}})\otimes I^{pol}_{2}.\label{S2}
\end{eqnarray}
Observe that the correlation can be written as
\begin{eqnarray}
C(\theta_{1},\phi_{1}, \theta_{2},\phi_{2})=(\Psi^{o}|(\sigma^{path}_{1\phi_{1},0}-\sigma^{path}_{1\phi_{1},\pi}).(\sigma^{pol}_{1\theta_{1},0}-\sigma^{pol}_{1\theta_{1},\pi})\\\otimes(\sigma^{path}_{2\phi_{2},0}-\sigma^{path}_{2\phi_{2},\pi}).(\sigma^{pol}_{2\theta_{2},0}-\sigma^{pol}_{2\theta_{2},\pi})|\Psi^{o}).
\end{eqnarray}
Consider the term
\begin{eqnarray}
(\Psi^{o} |\sigma^{path}_{1\phi_{1},0}.\sigma^{pol}_{1\theta_{1},0}\otimes\sigma^{path}_{2\phi_{2},0}.\sigma^{pol}_{2\theta_{2},0}| \Psi^{o}).\label{term1}
\end{eqnarray}
Let us define the intensity operators
\begin{eqnarray}
\hat{I}_{S_{1}}(\theta_{1},\phi_{1})=(\sigma^{path}_{1\phi_{1},0}.\sigma^{pol}_{1\theta_{1},0})\otimes \mathbb{I}_{S_{2}},\nonumber\\
\hat{I}_{S_{2}}(\theta_{2} ,\phi_{2})=\mathbb{I}_{S_{1}}\otimes(\sigma^{path}_{2\phi_{2},0}.\sigma^{pol}_{2\theta_{2},0}),\label{Intensity}
\end{eqnarray}
where $\mathbb{I}_{S_{1}}$ and $\mathbb{I}_{S_{2}}$ are identity operators.
Hence, the term (\ref{term1}) can be expressed as
\begin{eqnarray}
(\Psi^{o} |\sigma^{path}_{1\phi_{1},0}.\sigma^{pol}_{1\theta_{1},0}\otimes\sigma^{path}_{2\phi_{2},0}.\sigma^{pol}_{2\theta_{2},0}| \Psi^{o}) &=& (\Psi^{o} |\hat{I}_{S_{1}}(\theta_{1},\phi_{1})\hat{I}_{S_{2}}(\theta_{2},\phi_{2})|\Psi^{o})\nonumber\\&=&\frac{2|A_{1}|^{2}|A_{2}|^{2}}{(|A_{1}|^{2}+|A_{2}|^{2})^{2}}\left[1-\cos\left(\theta_{1}-\theta_{2}+\phi_{1} -\phi_{2}\right)\right]\label{HBTCORR}
\end{eqnarray}
which is the intensity-intensity correlation between $S_{1}$ and $S_{2}$ with the path phase $\phi_{1}$ and the polarization phase $\theta_{1}$ for source $S_{1}$ and the path phase $\phi_{2}$ and the polarization phase $\theta_{2}$ for $S_{2}$.
One can verify that all the sixteen terms of the correlation (\ref{cor}) can be expressed in a similar manner as intensity-intensity correlations, and that
the correlation function $C$ can be expressed compactly as 
\begin{eqnarray}
C(\theta_{1},\phi_{1},\theta_{2},\phi_{2})&=&\sum_{(k,l,m,n)=0}^{1}(-1)^{k+l+m+n}(\Psi^{o} |I_{S_{1}}(\theta_{1}+k\pi ,\phi_{1}+l\pi).\nonumber\\&.& I_{S_{2}}(\theta_{2}+m\pi,\phi_{2}+n\pi)|\Psi^{o})\nonumber\\
&=&\frac{4|A_{1}|^{2}|A_{2}|^{2}}{(|A_{1}|^{2}+|A_{2}|^{2})^{2}}\cos(\theta_{1}-\theta_{2}+\phi_{1}-\phi_{2}).
\end{eqnarray}

Note that
\begin{eqnarray}
(\Psi^{o} |\hat{I}_{S_{1}}(\theta_{1},\phi_{1})\hat{I}_{S_{2}}(\theta_{2},\phi_{2})|\Psi^{o})&=&(\Psi^{o}|\sigma^{path}_{1\phi_{1},0}.\sigma^{pol}_{1\theta_{1},0}\otimes\sigma^{path}_{2\phi_{2},0}.\sigma^{pol}_{2\theta_{2},0}|\Psi^{o})\nonumber\\ &=& (\Psi |\sigma^{path}_{10,0}.\sigma^{pol}_{10,0}\otimes\sigma^{path}_{20,0}.\sigma^{pol}_{20,0}|\Psi)\nonumber\\
&=&(\Psi |\hat{I}_{S_{1}}(0,0)\hat{I}_{S_{2}}(0,0)|\Psi)
\end{eqnarray}
where $|\Psi)$ is the symmetrized state (\ref{prestate}) before the beam splitter $BS^{'}$.
Using Eqn. (\ref{poststate}) observe that 
\begin{eqnarray}
(\Psi |\hat{I}_{S_{1}}(0,0)\hat{I}_{S_{2}}(0,0)|\Psi)&=&(\Psi ^f|\hat{U}_{BS^{'}}^{S_{1}\dagger}\sigma^{path}_{10,0}\hat{U}_{BS^{'}}^{S_{1}}.\sigma^{pol}_{10,0}\otimes\hat{U}_{BS^{'}}^{S_{2}\dagger}\sigma^{path}_{20,0}\hat{U}_{BS^{'}}^{S_{2}}.\sigma^{pol}_{20,0}|\Psi ^f).
\end{eqnarray}
Now,
 \begin{eqnarray}
 \hat{U}_{BS^{'}}^{S_{1}\dagger}\sigma^{path}_{10,0}\hat{U}_{BS^{'}}^{S_{1}}&=&|a)_{S_{1}}(a|_{S_{1}}\nonumber\\
 \hat{U}_{BS^{'}}^{S_{2}\dagger}\sigma^{path}_{20,0}\hat{U}_{BS^{'}}^{S_{2}}&=&|a)_{S_{2}}(a|_{S_{2}}\nonumber\\
 \end{eqnarray}
 Therefore the correlation (\ref{HBTCORR}) can be written in terms of the final state $|\Psi^{f})$ as
 \begin{eqnarray}
 (\Psi^{o} |\hat{I}_{S_{1}}(\theta_{1},\phi_{1})\hat{I}_{S_{2}}(\theta_{2},\phi_{2})|\Psi^{o})=
 (\Psi ^f|\left[|a)_{S_{1}}(a|_{S_{1}}.\sigma^{pol}_{10,0}\otimes |a)_{S_{2}}(a|_{S_{2}} .\sigma^{pol}_{20,0}\right]|\Psi ^f)
 \end{eqnarray}
 where, as can be seen from  (\ref{S1}) and (\ref{S2}),
 \begin{eqnarray}
 \sigma^{pol}_{10,0}=\mathbb{I}_{path}\otimes\frac{1}{2}\left[|H)_{S_{1}}+|V)_{S_{1}}\right]\left[(V|_{S_{1}}+(H|_{S_{1}}\right],\\
  \sigma^{pol}_{20,0}=\mathbb{I}_{path}\otimes\frac{1}{2}\left[|H)_{S_{2}}+|V)_{S_{2}}\right]\left[(V|_{S_{2}}+(H|_{S_{2}}\right].
 \end{eqnarray}
Hence,
\begin{eqnarray}
 (\Psi^{f} |\hat{I}_{S_{1}}(0,0)\hat{I}_{S_{2}}(0,0)|\Psi^{f})&=&(\Psi^{o} |\hat{I}_{S_{1}}(\theta_{1},\phi_{1})\hat{I}_{S_{2}}(\theta_{2},\phi_{2})|\Psi^{o}),
 \end{eqnarray}
and consequently,
\begin{eqnarray}
C(\theta_{1},\phi_{1},\theta_{2},\phi_{2})&=&\sum_{(k,l,m,n)=0}^{1}(-1)^{k+l+m+n}(\Psi^{f} |\hat{I}_{S_{1}}(k\pi ,l\pi). \hat{I}_{S_{2}}(m\pi,n\pi)|\Psi^{f})\nonumber\\
&=&\frac{2|A_{1}|^{2}|A_{2}|^{2}}{(|A_{1}|^{2}+|A_{2}|^{2})^{2}}\cos(\theta_{1}-\theta_{2}+\phi_{1}-\phi_{2}).
\end{eqnarray}
henceforth let us set $|A_{1}|^{2}=|A_{2}|^{2}$, i.e. the two sources to have the same intensity. Then
\begin{equation}
C=\cos\left(\theta_{1}-\theta_{2}+\phi_{1}-\phi_{2}\right).
\end{equation}
Note that the state (\ref{HBTcorr}) after the beam splitter $BS^{'}$ can be written in the form
\begin{eqnarray}
|\Psi^{f})&=&\frac{1}{\sqrt{2}}|a)_{S_{1}}|a)_{S_{2}}\left[|V)_{S_{1}}|V)_{S_{2}}-e^{i(\theta +\phi_{1}-\theta_{2}-\phi_{2})}|H)_{S_{1}}|H)_{S_{2}}\right]\nonumber\\&+&
\frac{1}{\sqrt{2}}|a)_{S_{1}}|b)_{S_{2}}\left[|V)_{S_{1}}|V)_{S_{2}}+e^{i(\theta +\phi_{1}-\theta_{2}-\phi_{2})}|H)_{S_{1}}|H)_{S_{2}}\right]\nonumber\\&+&\frac{1}{\sqrt{2}}|b)_{S_{1}}|a)_{S_{2}}\left[|V)_{S_{1}}|V)_{S_{2}}+e^{i(\theta +\phi_{1}-\theta_{2}-\phi_{2})}|H)_{S_{1}}|H)_{S_{2}}\right]\nonumber\\&+&\frac{1}{\sqrt{2}}|b)_{S_{1}}|b)_{S_{2}}\left[|V)_{S_{1}}|V)_{S_{2}}-e^{i(\theta +\phi_{1}-\theta_{2}-\phi_{2})}|H)_{S_{1}}|H)_{S_{2}}\right],
\end{eqnarray}
and hence the state along the path (a) from both the sources $S_1$ and $S_2$ is
\begin{eqnarray}
|\Psi^{f})_{aa}=\frac{1}{\sqrt{2}}|a)_{S_{1}}|a)_{S_{2}}\left[|V)_{S_{1}}|V)_{S_{2}}-e^{i(\theta +\phi_{1}-\theta_{2}-\phi_{2})}|H)_{S_{1}}|H)_{S_{2}}\right],
\end{eqnarray}
and its polarization part can be written in the form 
\begin{eqnarray}
|\Psi^{f})^{pol}_{aa}&=&\frac{1}{2}(1-e^{i(\theta_{1}+\phi_{1}-\theta_{2}-\phi_{2})})\frac{\left(|V)_{S_{1}}+|H)_{S_{1}}\right)}{\sqrt{2}}\frac{\left(|V)_{S_{2}}+|H)_{S_{2}}\right)}{\sqrt{2}}\nonumber\\ &+&\frac{1}{2}(1+e^{i(\theta_{1}+\phi_{1}-\theta_{2}-\phi_{2})})\frac{\left(|V)_{S_{1}}+|H)_{S_{1}}\right)}{\sqrt{2}}\frac{\left(|V)_{S_{2}}-|H)_{S_{2}}\right)}{\sqrt{2}}\nonumber\\&+&\frac{1}{2}(1+e^{i(\theta_{1}+\phi_{1}-\theta_{2}-\phi_{2})})\frac{\left(|V)_{S_{1}}-|H)_{S_{1}}\right)}{\sqrt{2}}\frac{\left(|V)_{S_{2}}+|H)_{S_{2}}\right)}{\sqrt{2}}\nonumber\\&+&\frac{1}{2}(1-e^{i(\theta_{1}+\phi_{1}-\theta_{2}-\phi_{2})})\frac{\left(|V)_{S_{1}}-|H)_{S_{1}}\right)}{\sqrt{2}}\frac{\left(|V)_{S_{2}}-|H)_{S_{2}}\right)}{\sqrt{2}}.
\end{eqnarray}
Note that all the terms have the same total phase difference $(\theta_{1}+\phi_{1}-\theta_{2}-\phi_{2})$, and contain in them the full phase information of the light from the two sources. The polarization filter $P_{45^{o}}$ projects the state to one of these components only, namely the component 
\begin{eqnarray}
|++)&=&\frac{|V)_{S_{1}}+|H)_{S_{1}}}{\sqrt{2}}.\frac{|V)_{S_{2}}+|H)_{S_{2}}}{\sqrt{2}}.
\een
Hence, what the detector D detects is the joint intensity of light given by 
\begin{eqnarray}
|(++|\Psi^{f})^{pol}_{aa}|^{2}=\frac{1}{2}\left(1-\cos(\theta_{1}+\phi_{1}-\theta_{2}-\phi_{2})\right).
\end{eqnarray}

\end{document}